\documentclass[10pt, conference]{IEEEtran}

\usepackage{xcolor}
\definecolor{ForestGreen}{HTML}{228B22}
\definecolor{cobaltblue}{RGB}{0,60,170}
\definecolor{darkbluegrey}{rgb}{0.164, 0.239, 0.400}

\usepackage{graphicx} %
\usepackage[margin=1in]{geometry}
\usepackage{subcaption}
\usepackage{amsmath}
\usepackage{amssymb}
\usepackage{subcaption}
\usepackage{graphicx}
\usepackage{caption}
\usepackage{svg}
\usepackage{tikz}
\usepackage{ifthen}

\usepackage[colorlinks=true, linkcolor=cobaltblue,citecolor=ForestGreen, urlcolor=darkbluegrey]{hyperref}
\usepackage{cleveref}

\usepackage{algorithm}
\usepackage{algpseudocode}
\usepackage{caption}
\usepackage{float}
\usepackage{placeins}
\usepackage{dblfloatfix}

\usepackage{changepage}
\usepackage{bbm}
\usepackage{physics}
\usepackage{xspace}

\crefname{equation}{Eq.}{Eqs.} 
\crefname{section}{Sec.}{Secs.} 
\crefname{figure}{Fig.}{Figs.} 
\crefname{table}{Table}{Tables} 
\crefname{appendix}{App.}{Apps.} 
\crefname{theorem}{Theorem}{Theorems}
\crefname{thm}{Theorem}{Theorems}
\crefname{defi}{Def.}{Defs.}
\crefname{conjecture}{Conjecture}{Conjectures}
\crefname{proposition}{Prop.}{Props.}
\crefname{lemma}{Lemma}{Lemmas}
\crefname{corollary}{Corollary}{Corollaries}

\usepackage{amsthm}
\newtheorem{theorem}{Theorem}
\newtheorem{corollary}{Corollary}
\newtheorem{definition}{Definition}

\usepackage{blindtext}

\newboolean{showcomments}
\setboolean{showcomments}{true}

\DeclareRobustCommand{\HL}[1]{\ifthenelse{\boolean{showcomments}}{{\textcolor{orange}{#1}}}{}}

\DeclareRobustCommand{\BA}[1]{\ifthenelse{\boolean{showcomments}}{{\textcolor{magenta}{#1}}}{}}

\newcommand{\hm}{\hspace{-0.15em}}
\newcommand{\bmt}{\boldsymbol{\theta}}
\newcommand{\peq}{\phantom{=}}

\newcommand{\dl}{\langle\langle}
\newcommand{\dr}{\rangle\rangle}

\newcommand{\Eb}{\mathbb{E}}
\newcommand{\Rb}{\mathbb{R}}
\newcommand{\Ib}{\mathbb{I}}
\newcommand{\Cb}{\mathbb{C}}
\newcommand{\Nb}{\mathbb{N}}

\newcommand{\Oc}{\mathcal{O}}

\newcommand{\Vc}{\mathcal{V}}
\newcommand{\Dc}{\mathcal{D}}
\newcommand{\Gc}{\mathcal{G}}

\newcommand{\lie}{\text{Lie}}

\newcommand{\SOc}{\mathcal{SO}}
\newcommand{\SUc}{\mathcal{SU}}

\newcommand{\so}{\mathfrak{so}}

\newcommand{\cry}{\mathtt{CR}^{\mathtt{Y}}}
\newcommand{\cy}{\mathtt{CY}}
\newcommand{\ry}{\mathtt{R}^{\mathtt{Y}}}
\newcommand{\py}{\mathtt{Y}}

\newcommand{\Gri}{\mathcal{G}^{\text{ring}}}
\newcommand{\Guni}{\mathcal{G}^{\text{uni}}}
\newcommand{\Gbeb}{\mathcal{G}^{\text{beb}}}

\newcommand{\GC}{\mathcal{G}^{[\cdot,\cdot]}}

\newcommand{\uri}{U^{\text{ring}}}
\newcommand{\uni}{U^{\text{uni}}}
\newcommand{\ubeb}{U^{\text{beb}}}

\newcommand{\I}{\mathbbm{1}}
\newcommand{\tvd}{\text{TVD}}

\newcommand\Tstrut{\rule{0pt}{2.6ex}}         %
\newcommand\Bstrut{\rule[-0.9ex]{0pt}{0pt}}   %

\title{Qvine: Vine Structured Quantum Circuits for Loading High Dimensional Distributions}

\author{
\IEEEauthorblockN{
David Quiroga\IEEEauthorrefmark{1}\IEEEauthorrefmark{2},
Hannes Leipold\IEEEauthorrefmark{1},
Bibhas Adhikari\IEEEauthorrefmark{1}
}
\IEEEauthorblockA{\IEEEauthorrefmark{1}\textit{Fujitsu Research of America} \\
Santa Clara, California, USA \\
}
\IEEEauthorblockA{\IEEEauthorrefmark{2}\textit{Department of Computer Science, Rice University}\\
Houston, Texas, USA \\ 
}
}

\date{\today}

\begin{document}

\raggedbottom

\maketitle
\begin{abstract}
Loading high dimensional distributions is an important task for utilizing quantum computers on applications ranging from machine learning to finance. The high dimensionality leads to a curse of dimensionality, representing a d-dimensional distribution with k resolution requires dk qubits and an unstructured parameterized circuit would express a unitary in an exponential operator space in the number of qubits, leading to vanishing gradients and poor convergence guarantees even at high depth. 

Vine copula decompositions are widely used to represent high dimensional distributions classically, showing high quality approximation in many important applications, such as financial modeling. We present Qvine, a vine structured ansatz for quantum circuits, that mirrors the vine decomposition to construct scalable quantum circuits with efficient trainability while achieving similarly high quality approximation for amplitude encoding distributions. For regular vines (R-vines), we show that the circuit depth scales at most quadratic in the dimension of the distribution, while for D-vines, as well as many practical R-vines, the circuit depth scales linear in the dimension. For 3-dimensional and 4-dimensional Gaussians and empirical joint stock price return distributions for selected stocks, our experiments show Qvines achieve high quality loading. 
\end{abstract}

\section{Introduction}\label{sec:intro}

Loading high-dimensional distributions remains a central task for realizing the utility of quantum computation in practical applications. Given that quantum amplitudes can encode a $2^n$-dimensional vector using only $n$ qubits, efficient data loading has become a crucial bottleneck in quantum algorithms, since the cost of exact state preparation could otherwise dominate the full computation and erase any downstream advantage~\cite{zoufal2019quantum, gonzalez2024efficient}. This concern is especially relevant for quantum workflows like quantum Monte Carlo, where a probability distribution is first prepared as a quantum state and then quantum algorithms like amplitude estimation~\cite{brassard2000quantum} are deployed to estimate expectations, tail risks, or payoffs, as in quantum risk analysis and option pricing~\cite{woerner2019quantum,stamatopoulos2020option}. %

As such, scalable quantum algorithms for encoding high dimensional distributions have become a major focus for the quantum academic and industrial community. Prior work includes qGANs~\cite{zoufal2019quantum} and Wasserstein qGANs~\cite{fuchs2023hybrid}, quantum multi-task learning~\cite{mourya2026contextual}, and hierarchical learning for quantum circuit Born machines (qCBM)~\cite{gharibyan2023hierarchical}.

Quality representation of high dimensional distributions is a heavily studied area in the classical literature as well. Vine copulas have been used successfully to represent high dimensional distributions in many domains including portfolio risk management, systemic risk, value at risk forecasting, and hydrological drought prediction~\cite{brechmann2013risk,pourkhanali2016measuring,zhang2014forecasting,wu2022predicting,low2018canonical}. Inspired by their efficiency and scalability in the classical literature, we propose a vine copula structured quantum circuit ansatz for loading high dimensional distributions. Copula-based loading of multivariate distributions into quantum circuits has been explored in Ref.~\cite{zhu2022generative}, where a copula is encoded directly into a maximally entangled quantum state. In contrast, our approach focuses on encoding a vine copula through a quantum circuit.

We find a vine copula decomposition for the $d$-dimensional distribution and use this structure to define a quantum circuit architecture (Qvines) for efficient, scalable distribution loading. With $k$ resolution for each dimension, our construction has quadratic scaling in the dimension $\Oc\hm\left( d^2 \, k \, L_b \right)$ for $R$-vine copulas while for $D$-vine copulas, it achieves linear scaling in the dimension $\Oc\hm\left( d \, k \, L_b \right)$ due to efficient compiling, where $L_b$ the number of layers for circuit blocks over $2k$ qubits each. With $L_b = \Oc\hm\left( 8^k \right)$, such parameterized circuit blocks have strong guarantees for any distribution~\cite{larocca2023theory}, while for realistic distributions the scaling can be significantly more modest. Note that typically resolution $k$ is taken to grow only logarithmically in the dimension $\Oc\hm\left( \log\left( d / \epsilon \right) \right)$ for error tolerance $\epsilon$. 

We present a progressive training methodology that has provable training guarantees by training each bivariate entangling block based on the order of its appearance in the vine. Our numerical simulations support that the resulting circuit can approximately load distributions with high accuracy in the case of $3D$ and $4D$ Gaussian distributions as well as the empirical 3-asset and 4-asset log-return distributions for selected technology stocks from the S$\&$P500 as well as the S$\&$P500 market index itself~\cite{fmp_sp500}. 

\section{Preliminaries}\label{sec:prelim}

\subsection{Copula Decomposition of Distributions}

Modeling dependence among multiple random variables is a fundamental problem in multivariate statistics, econometrics, and financial risk management~\cite{aas2016pair}. 
Traditional approaches often rely on covariance or correlation matrices such as the Pearson product-moment correlations, which captures only linear dependence. 
Copula theory provides a powerful framework for separating marginal distributions from the dependence structure of multivariate random variables. 
As such, vine copulas are widely used in financial econometrics to model nonlinear dependence and tail dependence between asset returns~\cite{dissmann2010statistical}.

A copula is a multivariate distribution defined on the unit hypercube that links marginal distributions to their joint distribution. 
A $d$-dimensional copula is a multivariate cumulative distribution function $C:[0,1]^d \to [0,1]$ with uniformly distributed marginals. 
The corresponding copula density for an absolutely continuous copula can be obtained as 
\begin{align} 
c(u_1,\hdots,u_d)=\dfrac{\partial^d}{\partial u_1\cdots\partial u_d} C(u_1,\hdots,u_d)
\end{align}
for all $u=(u_1,\hdots,u_d)\in [0,\, 1]^d$~\cite{czado2019analyzing}, see also Ref.~\cite{nelsen2006introduction}. A key result underlying copula theory is Sklar's theorem~\cite{Sklar1959}.

\begin{theorem}[Sklar's Theorem]
Let $F$ be a joint cumulative distribution function with univariate marginal distributions $F_1,\ldots,F_d$. Then there exists a copula $C$ such that
\begin{align} 
F(x)= C\big(F_1(x_1),\ldots,F_d(x_d)\big)
\end{align}
with associated density function 
\begin{align}
f(x)=c(F_1(x_1),\ldots, F_d(x_d)) \, \prod_{j=1}^{d} f_j(x_j) .
\end{align} 
For absolutely continuous distributions, the copula $C$ is unique.
\end{theorem}
The converse of Sklar's Theorem is also true and so the dependency structure among random variables is captured entirely by the copula.

\subsection{Vine Copula Decompositions for Multi-Dimensional Distributions}

While several parametric copulas exist, such as the class of Archimedean copulas for low-dimensional problems, specifying flexible copulas in high dimensions is challenging. 
Vine copulas address this limitation by decomposing a multivariate copula into a sequence of bivariate copulas arranged in a graphical structure known as a vine~\cite{bedford2002vines}. 
Let $X = (X_1,\dots,X_d)$ such that each $X_{j}$ is a random variable with univariate marginal distributions $F_j$. Define $u_i = F_i(X_i)$ so that $u_i \sim \text{Uniform}(0,1)$. 
The joint density can be expressed using a pair-copula construction (PCC) defined by a sequence of $d-1$ trees $T_1,\hdots, T_{d-1},$ called the vine:
\begin{align}
f(x) = \left( \prod_{\ell=1}^{d-1} \prod_{e\in T_\ell} c_{i_e,j_e;D_e} \left( u_{i_e|D_e},u_{j_e|D_e} \right) \right) \prod_{i=1}^{d} f_i(x_i),
\end{align}
where
\begin{itemize}
\item $T_\ell$ denotes the $\ell$-th tree in the vine,
\item $c_{i,j;D}$ denotes a conditional pair-copula density,
\item $D$ denotes the conditioning set associated with edge $e$.
\end{itemize}

For instance, for $d=3$ a pair copula construction of a joint parametric density is given by
\begin{eqnarray}
f(x_1,x_2,x_3;\bmt) &=& c_{13;2}(F_{1|2}(x_1|x_2), F_{3|2}(x_3|x_2); \theta_{13;2}) \nonumber \\ && \times c_{23}(F_2(x_2),F_3(x_3);\theta_{23})\nonumber \\
&& \times c_{12}(F_1(x_1),F_2(x_2); \theta_{12}) \nonumber \\
&& \times f_3(x_3) \, f_2(x_2) \, f_1(x_1), \nonumber
\end{eqnarray} where $c_{13;2}(\cdot,\cdot;\theta_{13;2})$, $c_{12}(\cdot,\cdot;\theta_{12}),$ $c_{23}(\cdot,\cdot;\theta_{23})$ are arbitrary parametric bivariate copula densities, and the parameter vector $\bmt=(\theta_{12},\theta_{23},\theta_{12;3})$~\cite{czado2019analyzing}. Consequently, this hierarchical decomposition allows flexible modeling of high-dimensional dependence structures. For further details, see Refs.~\cite{aas2009pair,kurowicka2006uncertainty}.

\begin{figure}[!t]
\centering
\includegraphics[width=0.5\textwidth]{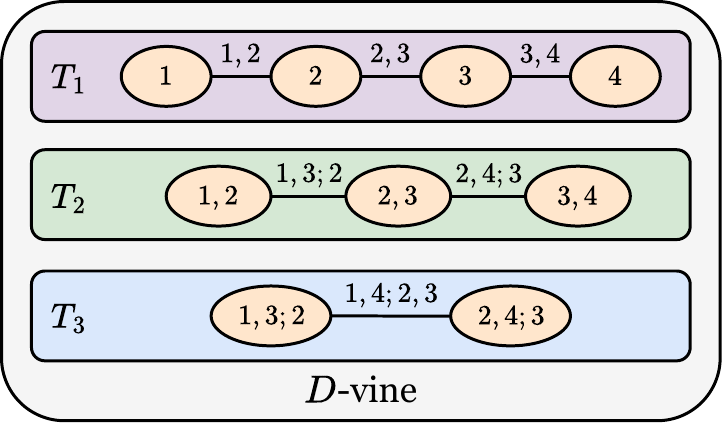}
\caption{\textbf{Example of a $4$-dimensional D-vine.}}
\label{fig:4ddvine}
\end{figure}

\begin{figure*}[!t]
\centering 
\includegraphics[trim=0cm 0 0.2cm 0, clip, width=1.0\textwidth]{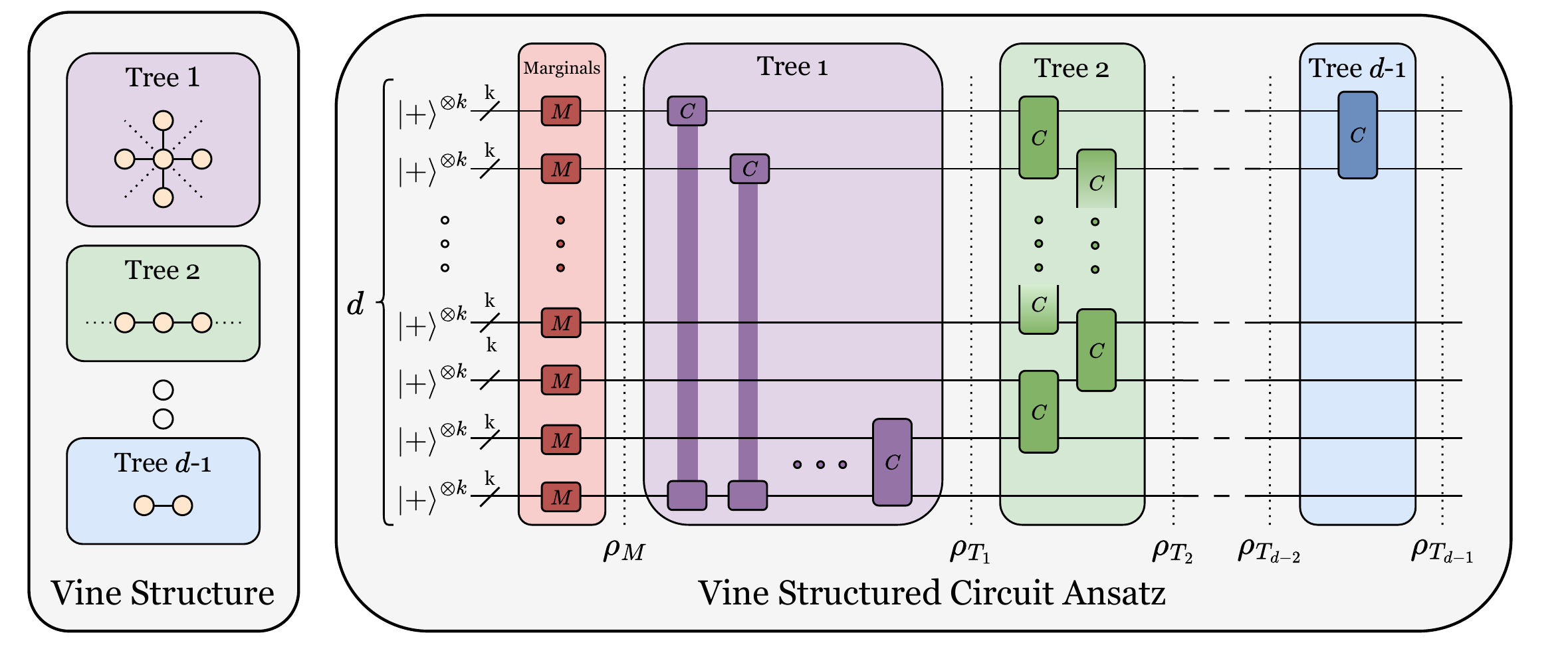}
\caption{\textbf{Qvines: Vine Structured Quantum Circuits.} Given a vine copula decomposition (depicted on the left), we define a vine structured quantum circuit (depicted on the right) based on the edges of each tree in the vine. Tree $1$ depicts a star graph while Tree $2$ is a path.}
\label{fig:vine_circ}
\end{figure*}

\subsubsection{Regular Vine (R-vine)}

A regular vine (R-vine) is a graphical model consisting of a sequence of linked trees
$T_1, T_2, \dots, T_{d-1},$
where each tree $T_k$ has nodes corresponding to the edges of the previous tree $T_{k-1}$. 
The trees must satisfy the proximity condition: two nodes in tree $T_k$ can be connected only if their corresponding edges in $T_{k-1}$ share a common variable.

The R-vine structure provides the most general framework for vine copulas, allowing arbitrary dependence patterns among variables. 
Each edge in the vine corresponds to a pair-copula that models conditional dependence between two variables given a conditioning set.

\subsubsection{Canonical Vine (C-vine)}

A canonical vine (C-vine) is a special case of an R-vine where each tree contains a central node connected to all other nodes. 

C-vines are suitable when one or a few variables exert dominant influence on the others. 
In financial applications, a market index or macroeconomic factor may serve as the root node of a C-vine structure~\cite{allen2013financial,brechmann2012truncated,allen2017risk}.

\subsubsection{Drawable Vine (D-vine)}

A drawable vine (D-vine) is another special case of an R-vine in which variables are arranged sequentially. 

D-vines are particularly useful for modeling ordered data such as time series, where dependence primarily occurs between neighboring observations~\cite{dissmann2010statistical, min2011bayesian, almeida2012modeling}.

The construction of a vine copula begins with selecting the edges of the first tree $T_1$, which determines the primary pairwise dependencies among the variables. 
In practice, the edges of $T_1$ are typically chosen based on empirical measures of dependence computed from the transformed uniform variables $u_i = F_i(x_i)$. 
A common approach is to compute pairwise dependence measures such as Kendall's $\tau$ or Spearman's $\rho$, and then construct a maximum spanning tree using these values as edge weights. 
The resulting tree captures the strongest pairwise dependencies in the first level of the vine.

\section{Vine Structured Quantum Circuits}\label{sec:vine_circ}

In this section, we delineate a quantum circuit architecture based on a vine structure such that the circuit can be employed to load high dimensional probability distributions on the quantum registers. 

To load the univariate distributions, we adapt the hierarchical quantum circuit architecture as developed in Ref.~\cite{gharibyan_hierarchical_2023}, which first captures the probability mass on the most significant qubits and subsequently refines the distribution on finer scales by incorporating each qubit of further precision. 

To then capture the interdependencies of the features, we utilize bivariate entangling blocks (BEBs) that are placed based on the vine copula decomposition, thereby performing a similar role to each pair bivariate copula. Specifically, every edge of every tree in the vine selects two feature registers over which the BEB is placed. \cref{fig:vine_circ} depicts our approach at a higher level.

As demonstrated in our numerical experiments, vine copulas provide a scalable decomposition for parameterized quantum circuits that load good quality high dimensional distributions in a flexible manner. 

\subsection{Probability distribution discretization}\label{subsec:discretize}

\begin{figure}[!t]
\centering 
\includegraphics[width=0.90\columnwidth]{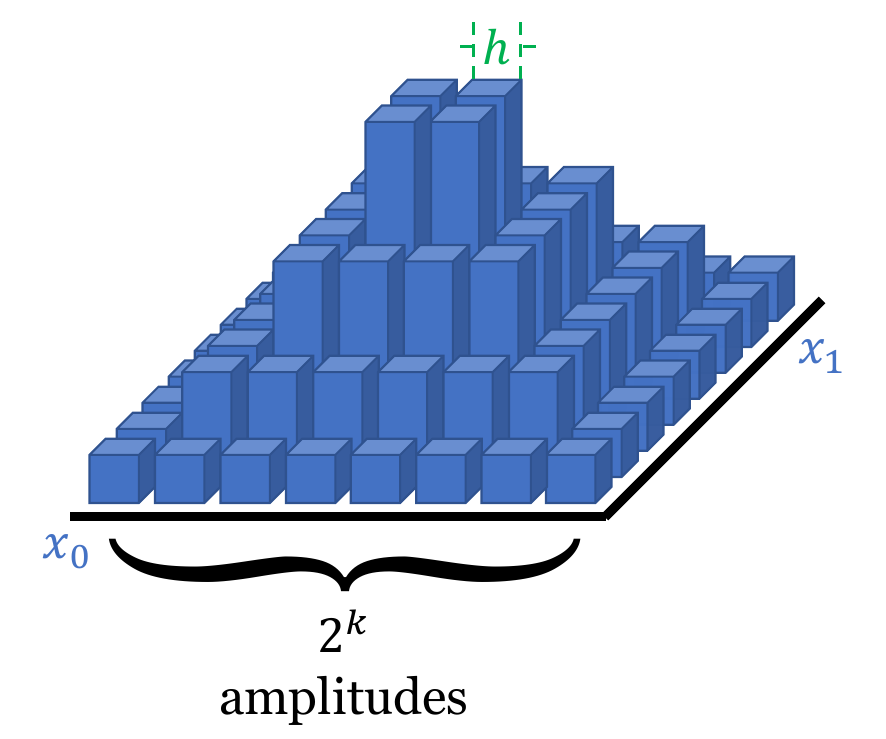}
\caption{\textbf{Discretization of a $2$-dimensional density function.} As discussed in \cref{subsec:discretize}, a continuous function $f$ over a domain $\Rb^{d}$ is well approximated as precision $k$ increases.}
\label{fig:discretization}
\end{figure}

We represent the joint density function $f(x)\equiv f(x_1,\ldots,x_d)$ of a $d$-dimensional continuous random vector $X=(X_1,\ldots,X_d)$ as a $2^{dk}$-dimensional quantum state by discretizing the domain of each random variable $X_j$ into $n=2^k$ bins (intervals) $\Ib_{j,l}$ for all $j$ with $1\leq l\leq n$. Each bin is indexed by a $k$-bit string in $\{0,1\}^k$. We then define a function $f_y$ that represents the probability mass (i.e. the volume of $f$) over the hyper-rectangular region corresponding to $y=(y_1,\ldots,y_d)$, where each $y_j \in \{0,1\}^{k}$ indexes a bin of $X_j$, for $1\leq j\leq d$. The $L_1$-error associated with approximating a given multivariate density $f$ using the aforementioned sample-based discretization scheme has been rigorously analyzed in Ref.~\cite{beirlant1998lrerror}. For instance, \cref{fig:discretization} describes the approximation function for $d=2$, where $h$ denotes the a fixed length of the discretizing interval.

We wish to find a quantum circuit to approximately encode the discretized distribution into a quantum state as follows:
\begin{equation}
\ket{f} = \sum_{y\in\{0,1\}^{dk}} \sqrt{f_y} \, \ket{y}.
\end{equation} 
Since $\ket{f}$ is normalized, it satisfies the total probability condition $ |\braket{f}{f} |^2 = \sum_{y\in\{0,1\}^{dk}} f_y = 1 $.

\subsection{Quantum Rotation and Control-Rotation Gates}\label{subsec:qgates}

Recall that the unitary of any rotation gate is described through complex exponentiation of its generator and its parameter $\theta$, $\mathtt{R^{g}}(\theta)= e^{-i \, \theta \, \mathtt{g} / 2} $, where $\mathtt{g}$ is the generator of the gate $\mathtt{R^{g}}$. Let $\delta_{jk}$ denote the Kronecker delta and recall that $ A^{0} = \I_{2} = \ketbra{0}{0} + \ketbra{1}{1} $ for any single-qubit operator $A \in \Cb^{2 \times 2}$. Then define the $j$ qubit standard basis embedded into the $n$-qubit operator space $\Cb^{2^{n} \times 2^{n}} $ as
\begin{align}
\ketbra{0}{0}_{j} &= \bigotimes_{k=1}^{n} \left( \ketbra{0}{0} \right)^{\delta_{jk}} \nonumber \\ 
&= \underbrace{\I_{2} \otimes \cdots \otimes \I_{2} }_{1:j-1} \otimes \ketbra{0}{0} \otimes \underbrace{\I_{2} \otimes \cdots \otimes \I_{2}}_{j+1:n} , \nonumber \\ 
\ketbra{0}{1}_{j} &= \bigotimes_{k=1}^{n} \left( \ketbra{0}{1} \right)^{\delta_{jk}} , 
\ketbra{1}{0}_{j} = \bigotimes_{k=1}^{n} \left( \ketbra{1}{0} \right)^{\delta_{jk}} , \nonumber \\ 
\ketbra{1}{1}_{j} &= \bigotimes_{k=1}^{n} \left( \ketbra{1}{1} \right)^{\delta_{jk}} ,\peq \I = \bigotimes_{k=1}^{n} \I_{2} . 
\end{align}

Then we can define the Pauli-Y operator acting on qubit index $j$ as $\py_{j} = -i \ketbra{0}{1}_{j} + i\ketbra{1}{0}_{j}$ and the parameterized $\py$ rotation gate that $\py_{j}$ generates as 
\begin{align}
\ry_{j}(\theta) &= e^{-i \, \theta \, \py_{j} / 2} = \cos(\theta/2) \, \I - i \sin(\theta/2) \, \py_{j} \nonumber \\ 
&= \cos(\theta/2) \ketbra{0}{0}_{j} - \sin(\theta/2) \ketbra{0}{1}_{j} \nonumber \\ 
&\peq + \sin(\theta/2) \ketbra{1}{0}_{j} + \cos(\theta/2) \ketbra{1}{1}_{j} . 
\end{align}

Then the control-$\py$ rotation gate, with qubit index $j$ as the control and index $k$ as the target, is generated by $ \cy_{j,k} = \ketbra{1}{1}_{j} \py_{k} $ as
\begin{align}
\cry_{j,k}(\theta) &= e^{-i \, \theta \,  \cy_{j,k} / 2} = e^{-i \, \theta \, \ketbra{1}{1}_{j} \py_{k}/2} \nonumber \\  
&=\left( \ketbra{0}{0}_{j} \I \right) + \left( \ketbra{1}{1}_{j} \, \ry_{k}(\theta) \right) .
\end{align}

Note that $\I, -i \, \py, -i \, \cy, \ry, \cry \in \Rb^{2^{n} \times 2^{n}} $.

\begin{figure}
\centering 
\includegraphics[trim=1.2cm 0cm 0cm 0cm, clip, width=0.48\textwidth]{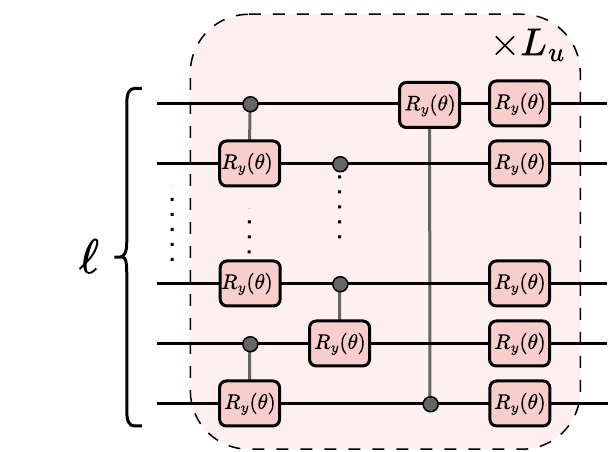}
\caption{\textbf{Special orthogonal group ring block.} Over $\ell$ qubits, by repeating this block structure $L_{u}$ times, the circuit can express unitaries in $\SOc\hm\left(2^{\ell}\right)$. }
\label{fig:circ_ring}
\end{figure}

\subsection{Special Orthogonal Group Ring Block}\label{subsec:ring_circ}

Define integer intervals $a:b = \{ a, a+1, \ldots, b - 1 \} $ inclusive of $a$ but exclusive to $b$. We have possible feature registers $r \in \{ 1, \ldots, d \} $, each with $k$ qubits. Then let $r_1 = k(r-1) + 1$ label the first qubit of each feature register. The basic building block of our circuit architecture is the Special Orthogonal Group $\SOc$ ring block (SORB), defined through the (ordered) generator set
\begin{align}
\Gri_{r_1:r_1+\kappa} &= \{ \cy_{r_1+j,r_1+j+1} \}_{j=0}^{\kappa-2} \cup \{ \cy_{r_1+\kappa-1,r_1} \} \nonumber \\ 
&\peq \cup \{ \py_{r_1+j} \}_{j=0}^{\kappa-1} . 
\end{align} 

The block has an associated dynamic Lie algebra~\cite{zeier2011symmetry, larocca2022diagnosing} that is shown to be isomorphic to $\so\hm\left( 2^{\kappa} \right)$. Recall that $\SOc\hm\left( N \right)$ is the subgroup of those unitaries in $\SUc\hm\left( N \right)$ keep the real subspace $\Rb^{N} \subset \Cb^{N}$ invariant.

Given angle parameters $\bmt \in \left[ -\pi,\pi \right]^{2 \kappa}$, the associated unitary for a single SORB layer is 
\begin{align}
\uri_{r_1:r_1+\kappa} \left( \bmt \right) &= \left( \prod_{j=0}^{\kappa-1} \ry_{r_1+j} \left( \theta_{j+\kappa+1} \right) \right) \cry_{r_1+\kappa-1,r_1}(\theta_{\kappa}) \nonumber \\ 
&\peq \times \left( \prod_{j=0}^{\kappa-2} \cry_{r_1+j,r_1+j+1}\left( \theta_{j+1} \right) \right)  ,
\end{align}

with $\uri$ compiled as shown in \cref{fig:circ_ring} to apply even and then odd pairings of $\cry$ followed by $\ry$. Then, for $L$ layers and $\bmt \in \left[ -\pi, \pi \right]^{2 \kappa L}$, repeating this block structure $L$ times defines 
\begin{align} 
\uri_{r_1:r_1+\kappa}\left( \bmt; L \right) = \prod_{\ell=1}^{L} \uri_{r_1:r_1+\kappa}\left( \bmt_{2(\ell-1)\kappa+1 : 2\ell \kappa} \right) ,
\end{align}
which is depicted visually in \cref{fig:circ_ring}. For example, if $k=3$ and $d=2$, then $r \in \{ 1, 4 \} $ and then a generator ring over the first register is 
\begin{align} 
\Gri_{1:4} = \{ \cry_{1,2}, \cry_{2,3}, \cry_{3,1} \} \cup \{ \py_{1}, \py_{2}, \py_{3} \} , 
\end{align} 
and over the second register is 
\begin{align} 
\Gri_{4:7} = \{ \cry_{4,5}, \cry_{5,6}, \cry_{6,4} \} \cup \{ \py_{4}, \py_{5}, \py_{6} \} .
\end{align}

\begin{figure}[!t]
\centering 
\includegraphics[trim=0.2cm 0 0cm 0, clip, width=0.50\textwidth]{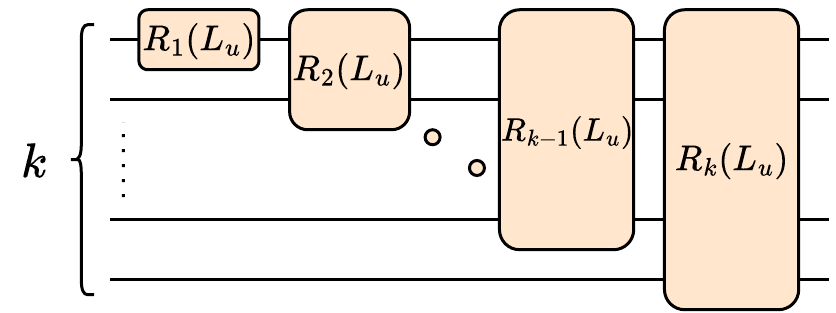}
\caption{\textbf{Hierarchical circuit for loading univariate distributions.} Over the $k$ qubits of each feature register, we employ this hierarchical circuit based on sequential blocks as shown in \cref{fig:circ_ring} with increasing number of qubits.}
\label{fig:circ_univar}
\end{figure}

Through the Lie correspondence, the space of operators which a parameterized circuit can express at any depth is described by the Lie algebra given by the closure of nested commutation of the generators of that circuit. Formally, we can define a commutator hierarchy starting with the generators $\GC_{1} = \Gc $ and level $k$ commutator set given by commuting the level $k-1$ set with the generators 
\begin{align} 
\GC_{k} = \left\{ [ G_{j} , H_{\ell} ] : G_{j} \in \Gc, H_{\ell} \in \GC_{k-1} \right\} . 
\end{align} 

At sufficiently high depth $L$, the gate set is sufficient to express unitaries in the Dynamic Lie Algebra (DLA) of the generator set. 
\begin{definition}[Dynamic Lie Algebra]
\begin{align} 
\langle \Gc \rangle_{\lie} &= \text{span}_{\Rb} \left(  \bigcup_{k=1}^{\infty} \GC_{k} \right) .
\end{align}
\end{definition}

The DLA of the SORB is the Lie algebra of the special orthogonal group~\cite{wiersema2024classification}\footnote{Associated with the $1$ dimensional periodic boundary Kitaev model with a longitudinal field.}.
\begin{theorem}\label{thm:soring}(SORB generates SO)
\begin{align}
\langle \Gri_{r_1:r_1+\kappa} \rangle_{\lie} \cong \so\hm\left( 2^{\kappa} \right). %
\end{align}
\end{theorem}

\begin{figure}[!t]
\centering 
\includegraphics[trim=0.2cm 0 0.0cm 0, clip, width=0.50\textwidth]{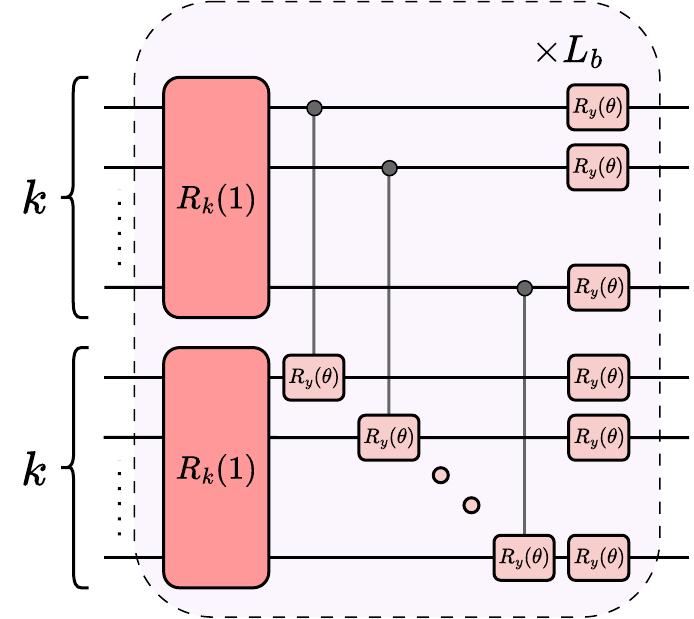}
\caption{\textbf{Bivariate entangling block.} Over two feature registers, each with $k$ qubits, this block structure repeated $L_{b}$ times captures bivariate dependency between the two features. }
\label{fig:circ_bivar}
\end{figure}

Because the DLA of the SORB generators is isomorphic to $\mathfrak{so}\hm\left( 2^{\kappa} \right)$, the family of unitaries generatable by increasing circuit depth $L$ through parameters $\bmt \in \left[ -\pi,  \pi \right]^{2 \kappa L}$ is dense in $\SOc\hm\left( 2^{\kappa} \right)$, which means any special orthogonal operator on the $k$-qubit register can be approximated arbitrarily well.
\begin{corollary}
$\left\{ \uri_{r_1:r_1+\kappa}( \bmt; L) : L \in \Nb, \bmt \in \left[ -\pi, \pi \right]^{2 \kappa L} \right\} $ is dense in $\mathcal{SO}\hm\left( 2^{\kappa} \right)$. 
\end{corollary} 

In particular, at depth $\Oc\hm\left(4^{\kappa}\right)$~\cite{larocca2023theory} overparameterization occurs, which leads to high quality representation for almost any unitary since there is sufficient control through the parameters of the associated Lie algebra. However, for structured learning tasks such as loading realistic distributions, significantly lower depth can be sufficient. 

As a result, a real amplitude valued wavefunction $\ket{\psi} $ will remain totally real under evolution of a unitary inside $\mathcal{SO}\hm\left( 2^{n} \right)$. 

\begin{theorem}[Preservation of Realness under SO]
If $\ket{\psi} \in \Rb^{2^{n}} $ and $U \in \mathcal{SO}(2^{n}) $, then $U \ket{\psi} \in \Rb^{2^{n}} $.
\end{theorem}

Note that the unitary group isomorphic to $\mathcal{SO}\left( 2^{k} \right)$ for a SORB placed on any specific register is a subgroup of $\mathcal{SO}\left( 2^{n} \right)$.

\subsection{Hierarchical Circuits for Loading Univariate Distributions}\label{subsec:uni_circ}

\newcommand{\te}{\mathtt{tend}}
\newcommand{\algprogtrain}{\texttt{train\_through\_vine}$(\Vc, d, k, L_b)$}
\begin{figure}[!t]
\centering
\begin{minipage}{1.0\linewidth}
\begin{algorithm}[H]
\caption{\algprogtrain}\label{alg:progtrain}
\begin{algorithmic}[1]
\Statex \textbf{Inputs: $\Vc = \{ T_{1}, \ldots, T_{d-1} \} $}
\For{$T_{j} = (V_{j}, E_{j}) \in \Vc $}
    \For{$ e_{jk} = (S_{v}, S_{w}) \in E_{j} $} 
        \State $\{x,y\} = (S_{v} \cup S_{w}) \setminus ( S_{v} \cap S_{w} ) $
        \State $\bmt^{n} \sim U_{2 k L_b}(-a/2,a/2)\text{, }a\ll 1 $ \State $ \ket{\psi} \leftarrow U_{\Vc}|s\rangle$
        \State Train $\ubeb_{x,y}\left( \bmt^{n} \right) \ket{\psi}$ 
        \State $U_{\Vc} \leftarrow U_{\Vc} \, \ubeb_{x,y}$ \Comment{extend circuit to include new block.}
        \State $\bmt \leftarrow \bmt \oplus \bmt^{n}$ \Comment{append new parameters.}
    \EndFor
\EndFor
\State \textbf{return } $U_{\Vc}(\bmt)$
\end{algorithmic}
\end{algorithm}
\end{minipage}
\caption*{Alg. \ref{alg:progtrain}: \textbf{Progressive training through the vine.} For every tree $T_{j} = (V_{j}, E_{j}) $ in the input vine $\Vc$, we consider each edge $(S_{v}, S_{w}) \in E_{j}$ that specifies two features $x,y$ through the symmetric difference. We train entangling bivariate layer (see Fig.~\ref{fig:circ_bivar}) on $x,y$ and append this to the vine circuit. }
\end{figure}

In Ref.~\cite{gharibyan2023hierarchical}, the authors proposed a hierarchical structure-based quantum circuit for loading probability distributions into the squared amplitudes of computational basis vectors represented by bitstrings, adapting the Quantum Circuit Born Machines framework. The hierarchical ansatz can be viewed as a sequence of nested variational circuits that are progressively expanded during the training process. 

We describe a similar hierarchical circuit for each univariate distribution based on SORB that begins with the most significant bits, learns a circuit structure to approximately represent this bit and then progressively learns to approximately represent up to $j$ significant bits until $j$ is equal to the desired resolution $k$ as discussed in \cref{subsec:discretize}.

Given register $r$, the first qubit is $r_{1} = k(r-1) + 1$. Then $L$ layers with parameters $\bmt \in \left[ -\pi, \pi \right]^{k(k+1) L}$, as depicted \cref{fig:circ_univar}, describes a unitary
\begin{align}
\uni_{r} \left( \bmt; L \right) = \prod_{j=1}^{k} U_{r_1:r_1+j}^{ring} \left( \bmt_{(j-1)L:jL} \right),
\end{align}

with the associated generator set
\begin{align}
\Guni_{r} = \bigcup_{j=1}^{k} \Gri_{r_1:r_1+j} ,
\end{align}

From \cref{thm:soring}, it immediately follows that this hierarchical structure with sufficient $L$ can express $\mathfrak{so}$ since $\Gri_{r:r+k} \subset \Guni_{r} $. 

\begin{theorem}(Hierarchical SORB generates SO)
$ \langle \Guni_{r} \rangle \cong \so\hm\left( 2^{k} \right) $ .
\end{theorem}

\subsection{Entangling Bivariate Layers based on Pair Copulas}\label{subsec:bi_circ}

To entangle two feature registers, we define the bivariate entangling block (BEB) that uses a SORB on each feature register before applying $\cry$ gates across the two registers between each resolution bit followed by single $\ry$ gates on each qubit. \cref{fig:circ_bivar} depicts the circuit we describe and each $C$ labeled cross register block in \cref{fig:vine_circ} is such a BEB block.  

Given registers $r,q$, the first qubits are $r_1 = k(r-1)+1$ and $q_1 = k(q-1)+1$. Then over two registers $r$ and $q$, the generators of the BEB block are
\begin{align}
\Gbeb_{r,q} &= \{ \py_{r_1+j}, \py_{q_1+j} \}_{j=0}^{k-1} \cup \{ \cy_{r_1+j,q_1+j} \}_{j=0}^{k-1} \nonumber \\ 
&\peq \cup \Gri_{r_1:r_1+k} \cup \Gri_{q_1:q_1+k} .
\end{align}

At sufficient depth, arbitrary $\SOc$ rotations over the two registers is possible, signifying that the ansatz is well-suited to capture cross-correlations. Formally, as in \cref{subsec:ring_circ,subsec:uni_circ}, the DLA of the BEB generators is $\so$, now over two registers. 

\begin{theorem}(BEB generates SO)
$ \langle \Gbeb \rangle_\lie \cong \mathfrak{so}\hm\left( 2^{2k} \right) . $
\end{theorem}

Given parameters $\bmt \in [ -\pi, \pi ]^{7k}$, we define a single layer of the bivariate block ansatz as
\begin{align}
\ubeb_{r_1,q} \left( \bmt \right) &= \left( \prod_{j=0}^{k-1} \ry_{r_1+j}\left( \theta_{j+1} \right) \right) \left( \prod_{j=0}^{k-1} \ry_{q_1+j}\left( \theta_{k+j+1} \right) \right) \nonumber \\ 
& \times \left( \prod_{j=0}^{k-1} \cy_{r_1+j,q_1+j}\left( \theta_{2k+j+1} \right) \right) \nonumber \\ 
& \times \uri_{r_1,r_1:k} \left( \bmt_{3k+1:5k} \right) \uri_{q_1,q_1:k}\left( \bmt_{5k+1:7k} \right) 
\end{align}

And so for $L$ with parameters $\bmt \in \left[ -\pi, \pi \right]^{7kL} $, the unitary is $\ubeb_{r,q} \left( \bmt; L \right) = \prod_{\ell=1}^{L} \ubeb_{r,q} \left( \bmt_{(\ell-1)k+1:\ell k} \right) $.

\section{Progressively Training through the Vine}\label{sec:prog_train}

\begin{table}[!t]
\centering 
\begin{tabular}{l|l}
 & Vine ansatz with $d$-dimensions, resolution $k$, \Tstrut\\ 
 & $L_u$ univariate layers, $L_b$ bivariate layers \Bstrut\\ 
\hline
\# of trees in vine & $d-1$ \Tstrut\Bstrut\\ 
\# of edges in vine & $ d \, (d-1) / 2 $ \Tstrut\Bstrut\\ 
\# of nodes in vine & $ d \, (d+1) / 2 - 1  $ \Tstrut\Bstrut\\ 
\# of parameters & $ L_u \, (k+1) \, k \, d + 7 \, L_b \, k  \, d \, (d-1) / 2 $ \Tstrut\Bstrut\\ 
\# of $\ry$ gates & $  L_u \, k \, (k+1) \, d / 2  + 2 \, L_b \, k \, d \, (d-1) $ \Tstrut\Bstrut\\ 
\# of $\cry$ gates & $ L_u \, k \, (k+1) \, d \, / 2 + 3 \, L_b \, k \, d \, (d-1) / 2 $ \Tstrut\Bstrut 
\end{tabular}
\caption{\textbf{Resource Utilization of the Vine Ansatz Circuit.} Given a $d$-dimensional feature space with resolution $k$ for each feature, this table summarizes the resources of the VAC. Note that the number of parameters grows in $\Oc\left( d^2 \right)$ and that for a $D$-vine as well as many typical $R$-vines, the circuit depth grows in $\Theta\left( d \right)$.}
\label{tab:resources}
\end{table}

In this section, we describe how to progressively grow and train the Vine Structured Quantum Circuit in tandem. We describe Alg.~\ref{alg:progtrain} in words. Given an implicit $d$-dimensional density function $f(x_{1},\ldots,x_{d})$ defined over a dataset $\Dc = \{ (x_1^{j},\ldots,x_d^{j}) \}_{j=1}^{|\Dc|} $, we define $k$ qubit feature register for each feature $x_{j}$. Then the target wavefunction to approximate is
\begin{align}
\ket{ f } = \frac{1}{\sqrt{|\Dc|}} \sum_{(x_{1},\ldots,x_{d,1}) \in \Dc } \ket{ x_{1} } \ldots \ket{x_{d}} , 
\end{align} where $ x_{j} \in \{0,1\}^{k} $ with the standard bit to qubit mapping $ x_{j} \rightarrow \ket{ x_{j,1} } \ldots \ket{ x_{j,k} } $. This target could be altered to include regularization and data augmentation. 

\begin{figure*}
\centering 
\includegraphics[width=1.0\textwidth]{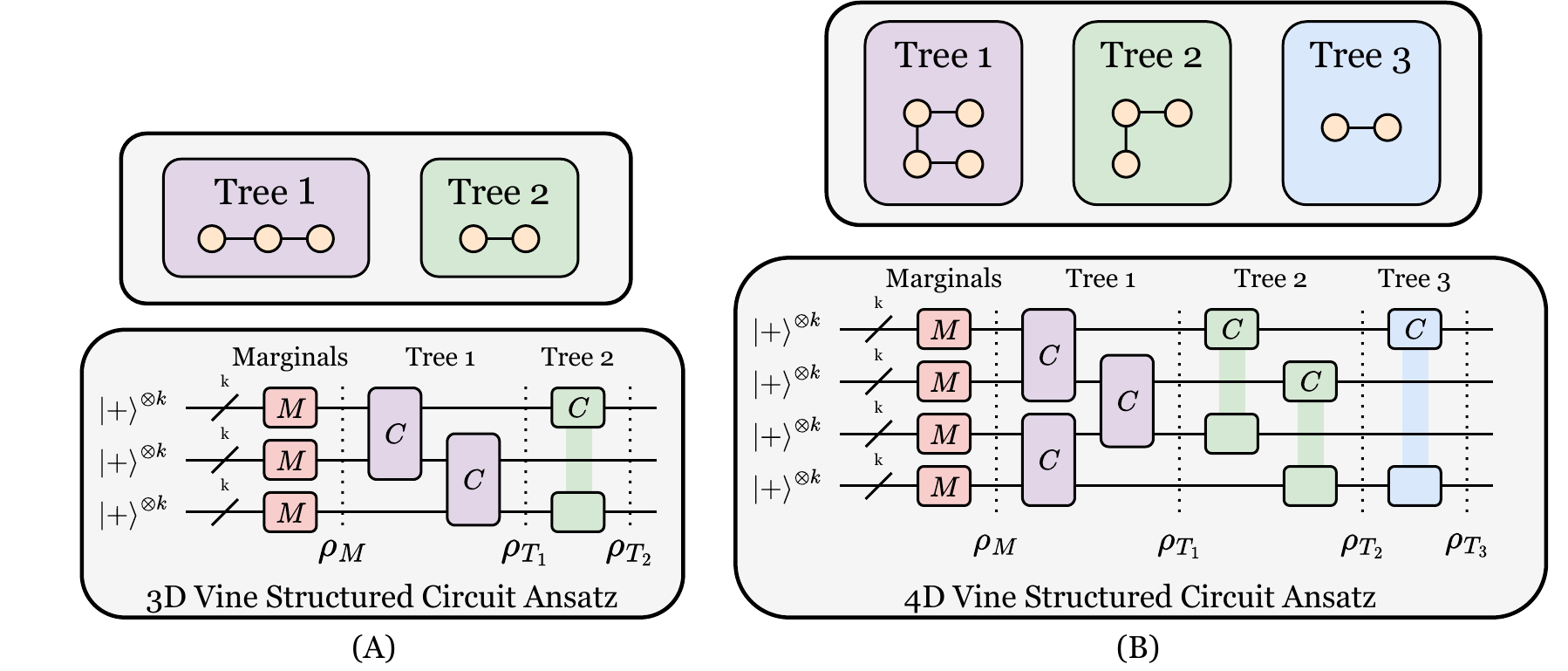}
\caption{\textbf{A Vine Structured Circuit for 3 \& 4 Dimensional Distributions.} In the case that $d=3$, we have a $3k$ qubit memory with precisely $2$ trees in the vine. In the case that $d=4$, we have a $4k$ qubit memory with precisely $3$ trees in the vine. (B) shows the $D$-vine we used in experiments, the vine is also depicted in \cref{fig:4ddvine}.}
\label{fig:vine_circ_34d}
\end{figure*}

We compute a vine copula decomposition $\Vc$ based on $\Dc$. We sequence the vine structure as $\Vc = \dl T_{1}, \ldots, T_{d-1} \dr$ with each tree given as a sequence of edges $ T_{j} = \dl e_{j,1}, \ldots, e_{j,d-j} \dr $ such that $ e_{j, k} = (S_{v}, S_{w}) $ for two sets of features $ S_{v}, S_{w} $. Due to the structure of the vine copula, $S_{v}$ corresponds to a set of features that are associated with one another through edges of the previous trees $T_{1}, \ldots, T_{j-1}$ such that \textit{two} features are \textit{uniquely} identified through the symmetric difference $ \{ x, y \} = (S_{v} \cup S_{w}) \setminus (S_{v} \cap S_{w}) $. We also use $T_{j,k}$ to refer to the $k$-th edge of the $j$-th tree.

We initialize the quantum system as the uniform superposition over all $k$ qubits of each of the $d$ feature registers
\begin{align}
\ket{ \psi_{0} } = \ket{+}^{\otimes dk} = \frac{1}{\sqrt{2^{dk}}} \sum_{x \in \{ 0, 1 \}^{d k}} \ket{x} .
\end{align}

We begin our circuit construction and training by approximately loading the univariate distributions of each feature. The associated wavefunction for each feature is given by
\begin{align}
\ket{ f_{r} } = \frac{1}{\sqrt{|\Dc_{r}|}} \sum_{(x_{r}^{j}, \ldots, x_{r+k-1}^{j}) \in \Dc_{r} } \ket{ x_{r}^{j}, \ldots, x_{r+k-1}^{j} } , 
\end{align}
where $\Dc_{r} = \{ (x_{r,1}^{j}, \ldots, x_{r,k}^{j}) : x_{r}^{j} \in \Dc \} $ is what the discretized dataset $\Dc$ provides over the feature $r$. Let $ \ket{\psi_{r} } \approx \ket{ f_{r} } $ be the loaded wavefunction for each feature, then the wavefunction after loading each univariate distribution is
\begin{align}
\ket{ \psi_{M} } = \bigotimes_{j=1}^{d} \ket{ \psi_{j} } ,
\end{align}

as depicted in \cref{fig:vine_circ} with $\rho_{M} = \ketbra{\psi_{M}}{\psi_{M}}$. As described in \cref{subsec:uni_circ}, over each feature register $r$ we train a hierarchical circuit $ \uni_r (\bmt^{r}; L_{u}) $ over $\Dc_{j}$ through a fidelity loss 
\begin{align}
\left| \bra{f} \left( \prod_{r=1}^{d} \uni_r(\bmt^{r}) \right) \ket{+}^{\otimes d k} \right|^2 . 
\end{align}
Each $\bmt^{r}$ is randomly initialized over $ [ -a/2, a/2 ]^{k (k-1) L_{u}} $, $a \ll 1$, and after training
\begin{align}
\ket{ \psi_{M} } = U_{M} \left( \bmt^{M} \right) \ket{ \psi_{0} } , 
\end{align}
with $ U_{M}(\bmt^{M}) = \prod_{r=1}^{d} \uni_r(\bmt^{r}) $ as depicted in red in \cref{fig:vine_circ}, such that the parameters of $U_{M}$ are expressed through concatenation $ \bmt^{M} = \bigoplus_{r=1}^{d} \bmt^{r} $ in the range $ \left[ -\pi, \pi \right]^{d k (k-1) L_{u}} $. 

Now, we construct and train each bivariate block as described by the vine decomposition. Let us begin with the first tree $T_{1} = \dl e_{1,1}, \ldots, e_{1,d-1} \dr $ over $ d $ vertices, each a singleton set. Let $ e_{1,1} = (\{ r \}, \{ q \}) $ for some specific $r, q$. %
Then we train a bivariate entangling block (BEB) unitary $ \ubeb_{r,q}(\bmt^{n}; L_{b}) $ with the fidelity loss
\begin{align} 
\left| \bra{ f } \ubeb_{r,q}(\bmt^{n}; L_{b}) \ket{\psi_{M}} \right|^2 , 
\end{align} 
or sample loss 
\begin{align}
\Eb_{x \in \Dc} \left[ \left| \bra{x} \ubeb_{r,q} \left( \bmt^n \right) \ket{\psi_{M}} \right|^2  \right]
\end{align}

$ \bmt^{n} $ is randomly initialized on $[ -a/2, a/2 ]^{3 k L_{b}}$. After training, we have
\begin{align}
\ket{ \psi_{e_{1,1}} } &= U_{e_{1,1}} \left( \bmt^{e_{1,1}} \right) \ket{\psi_{0}} \nonumber\\
&= \ubeb_{r,q} \left( \bmt^{n} \right) U_{M} \left( \bmt^{M} \right) \ket{\psi_{0}}, 
\end{align}
with $ \bmt^{e_{1,1}} = \bmt^{M} \oplus \bmt^{n} $. 

Similarly, for each edge $ e_{j,k} = (S_{v}, S_{w}) \in T_{j} $, we have $ \ket{\psi_{e_{j,k-1}}} = U_{e_{j,k-1}} \left( \bmt^{e_{j,k-1}} \right) \ket{\psi_{0}} $ and train a BEB unitary $\ubeb_{r,q}( \bmt^{n}; L) $ for $\{r,q\} = (S_{v} \cup S_{w}) \setminus (S_{v} \cap S_{w}) $. 

Then the unitary is trained through fidelity loss
\begin{align}
\left| \bra{f} \ubeb_{r,q} \left( \bmt^{n} \right)  \ket{\psi_{e_{j,k-1}}} \right|^2 , 
\end{align}
or sampling loss 
\begin{align}
\Eb_{x \in \Dc} \left[ \left| \bra{x} \ubeb \left( \bmt^{n} \right) \ket{\psi_{e_{j,k-1}}} \right|^2 \right] .
\end{align}

$ \bmt^{n} $ is randomly initialized on $[ -a/2, a/2 ]^{3 k L_{b}}$. After training, we have
\begin{align}
\ket{ \psi_{e_{j,k}} } &= U_{e_{j,k}} \left( \bmt^{e_{j,k}} \right) \ket{\psi_{0}} \nonumber\\
&= \ubeb_{r,q} \left( \bmt^{n} \right) U_{e_{j,k-1}} \left( \bmt^{e_{j,k-1}} \right) \ket{\psi_{0}},  
\end{align}
with $ \bmt^{e_{j,k}} = \bmt^{e_{j,k}} \oplus \bmt^{n} $. Note that we define $e_{j,0} = e_{j-1,d-j}$ when beginning on the first edge of a new tree. Let $\ket{\psi_{e_{j,d-j}}} = \ket{\psi_{T_{j}}}$ refer to the wavefunction after processing all edges of tree $T_{j}$, then $ \rho_{T_{j}} = \ketbra{\psi_{T_{j}}}{\psi_{T_{j}}}$ is depicted after each tree circuit in \cref{fig:vine_circ}.

\cref{tab:resources} summarizes the resource usage of our circuit architecture, given $L_u$, $L_b$, $d$, and $k$.

%
%
%
%
%

\section{Numerical simulations}\label{sec:numerics}

For all numerical simulations we train using the Adaptive Moment Estimation (ADAM) optimizer~\cite{kingma2014adam} with a patience parameter to reduce the learning rate when no improvements are made on the loss after a number of iterations in the Pennylane framework~\cite{bergholm2018pennylane}. We initialize all new parameters $\theta$ to a uniform distribution as $\theta \sim \text{Uniform}(-5\times 10^{-2},5\times 10^{-2})$ to perturb the loss function away from previously found local minima, and freeze previous parameters. We use discretized versions of probability distributions so that bins correspond to the amplitudes available on a circuit as in \cref{fig:discretization}. For a target density function $f(x)$ treated as a quantum state via $|\phi\rangle = \sum_{x} \sqrt{f(x)} \, \ket{x}$ and our generated distribution $|\psi\rangle$, we use the infidelity $1 - \mathcal{F}$ as the loss function of the generated state, with the fidelity given as
\begin{equation}
    \mathcal{F}=|\langle \psi|\phi\rangle|^2 = \left| \sum_{x \in \{ 0, 1 \}^{dk}} \sqrt{p(x)} \braket{\psi}{x} \right|^2 . 
\end{equation}

Our implementation represents the target distribution state as a density operator via $\rho_\phi = |\phi\rangle \langle \phi|$ to optimize over $\mathcal{F}=\langle \psi|\rho_\phi|\psi\rangle$. As a measure of the quality of loading, we utilize the L1 total variational distance for two discrete distributions $ \tvd = \frac{1}{2} \sum_{x \in \{0,1\}^{dk}} | f(x) - q_\Vc(x) |$ where $q_\Vc(x)$ is the probability distribution given by measuring the standard basis $q_\Vc(x) = | \braket{x}{\psi(\bmt_\Vc)} |^2 $ after training is completed. This fidelity formulation based on expectation measurements enables execution speedups when using GPU simulation. While our formulation reduces execution time by using a A100 GPU, we are still limited in the amount of CPU RAM available to construct $\rho_\phi \in \mathbb{R}^{2^{dk}\times 2^{dk}}$ as the operator to be measured, which we considered up to $dk = 12$. All our experiments use d-vines and Prim's algorithm~\cite{jarnik1930jistem, 6773228} for a greedy minimum spanning tree generation with Kendall's $\tau$ as the edge weight assignment criterion as set by default in the Python pyvinecopulib~\cite{nagler_2025_14841456} library. %

\subsection{Loading 3D and 4D Gaussian Distributions}

We train a QNN to replicate both a trivariate uncorrelated and a correlated Gaussian distribution using the vine structured circuit from \cref{fig:vine_circ_34d}(A), where each tree is trained progressively. The pdf of the uncorrelated Gaussian distribution is $f_{\text{un}}(x) \sim \mathcal{N}(\mu, \mathbf{\Sigma}_{\text{3D-uncor}})$ with $\mu= (0.05,0.05,0.05) $ and $\sigma_{\text{3D-uncor}} = 0.5 \, \I_{3}$. In contrast, the pdf of the correlated Gaussian distribution is $f_{\text{3D-cor}}(x) \sim \mathcal{N}(\mu, \mathbf{\Sigma}_{\text{3D-cor}})$, 
\begin{equation}
    \mathbf{\Sigma}_{\text{3D-cor}} = \begin{pmatrix} 0.05 & 0.03 & 0.015 \\
0.03 & 0.05 & -0.01 \\
0.015 & -0.01 & 0.05
\end{pmatrix}.
\end{equation}

\begin{figure}[!t]
\includegraphics[width=1\columnwidth]{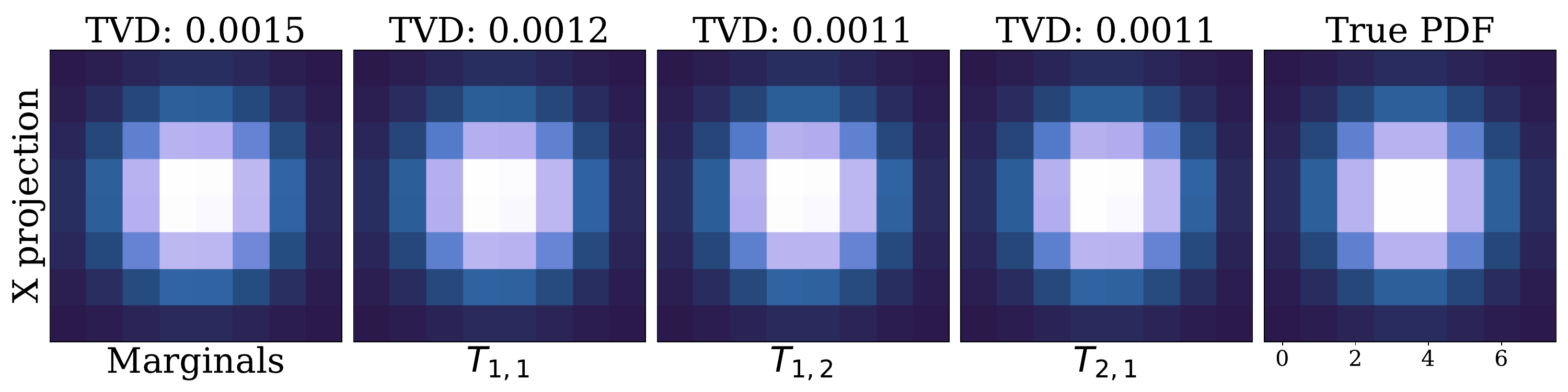}
\caption{\textbf{Progressive TVD of the trained QNN projected onto the X axis for the uncorrelated 3D Gaussian.} Each of the $3$ dimensions is discretized over $3$ qubits, with marginal loading as step $0$ and the next $3$ steps for each BEB of the $3$ edges in the vine. $T_{i,j}$ corresponds to the $j$-th edge of the $i$-th tree.}
\label{fig:gaussian_3D_iid_constant}
\end{figure}

For the uncorrelated Gaussian, we skip the intermediate hierarchical univariate layer to begin training the complete marginal distributions. Only one layer of our ansatz was enough to converge to an infidelity of $5\times10^{-4}$ and to a low TVD as shown in \cref{fig:gaussian_3D_iid_constant}.

The reason behind such quick convergence is that all the marginal distributions are equivalent and either enough layers for training each marginal independently or enough expressivity throughout all the layers to represent equal marginal distributions would suffice.

The generated correlated Gaussian distribution and its step-wise TVD can be visualized in \cref{fig:gaussian_3D_cubic}. We perform an ansatz ablation study on the number of layers for linear, quadratic and cubic scaling, with cubic scaling achieving the lowest TVD, as shown in \cref{fig:gaussian_3D_scaling_cubic}.

We train a QNN based on Alg.~\cref{alg:progtrain} via the circuit on \cref{fig:vine_circ_34d}(B) to load a 4-dimensional Gaussian distribution $f_{\text{4D-cor}}(x) \sim \mathcal{N}(\mu, \mathbf{\Sigma}_{\text{4D-cor}})$, with mean and standard deviation
\begin{align}
\mu &= \begin{pmatrix}
    0.05 & 0.05 & 0.05 & 0.05
\end{pmatrix} \text{, and} \\
\mathbf{\Sigma}_{\text{4D-cor}} &= \begin{pmatrix}
    0.05 & 0.03 & 0.015 & 0.01 \\
0.03 & 0.05 & -0.01 & 0.02 \\
0.015 & -0.01 & 0.05 & 0.025 \\
0.01 & 0.02 & 0.025 & 0.05
\end{pmatrix}.
\end{align}

The TVD between the loaded and actual distribution improves with every BEB block based on the vine, with a final $\tvd = 10^{-2}$. \cref{fig:gaussian_4D_fixed} depicts the first two dimensions of the loaded and actual distribution.

\subsection{Joint Log-Return Distributions of 3 and 4 stocks}

We employ our vine structured qauntum circuit to load the joint empirical probability distribution of asset price daily log-returns as variables: the S\&P500 index, the AMD stock ticker and the NVIDIA stock ticker to define $f_{\text{3D-ret}}$ as well as with the APPLE stock ticker to define $f_{\text{4D-ret}}$ with $3$ qubits per feature. Each empirical distribution is a discretization over the daily log-returns frequency distribution
\begin{align} 
\frac{1}{|\Dc|} \sum_{t \in [1, 1024]} \left(\log\left(\frac{p_1(t)}{p_1(t\!-\!1)}\right), \ldots, \log\left(\frac{p_k(t))}{p_k(t\!-\!1)} \right)\right) ,
\end{align}
with $p_{k}(t)$ as the end-of-day price of asset $k$ for normalized day $t \in [1, 1024]$ starting from 04-09-2022 to 04-09-2026, which we collected from the API of Ref.~\cite{fmp_sp500}.

\begin{figure*}[!t]
\centering 
\begin{subfigure}[t]{0.96\textwidth}
\includegraphics[width=0.96\textwidth]{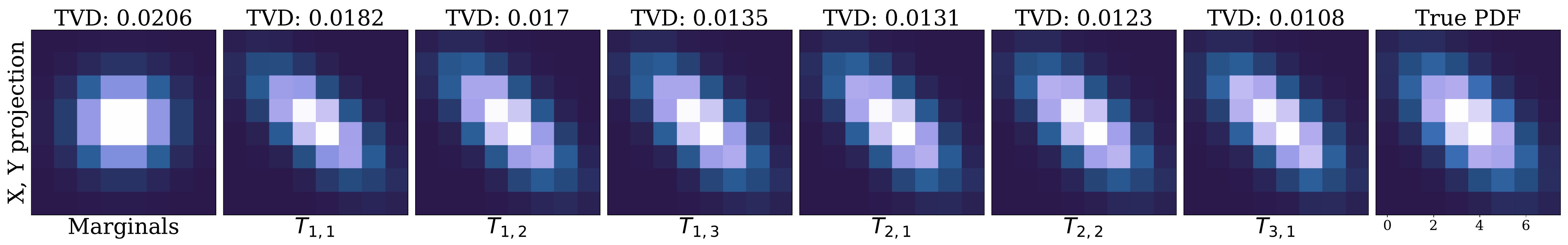}
\end{subfigure}
\caption{\textbf{TVD of the progressive trained QNN projected onto the X, Y axis for the 4D Gaussian.}}
\label{fig:gaussian_4D_fixed}

\vspace{0.5em}

\begin{subfigure}[t]{0.96\textwidth}
\includegraphics[width=0.96\textwidth]{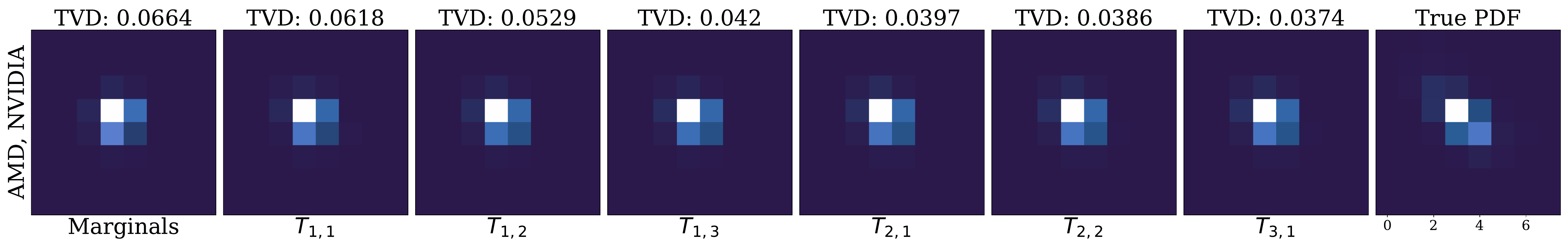}
\end{subfigure}
\caption{\textbf{TVD of the progressive trained QNN for the 4D log-return distribution on selected stocks.}}
\label{fig:stocks_4D_fixed}

\vspace{0.5em}

\begin{tabular}{c c}
\begin{subfigure}[t]{0.96\columnwidth}
\includegraphics[width=0.96\columnwidth]{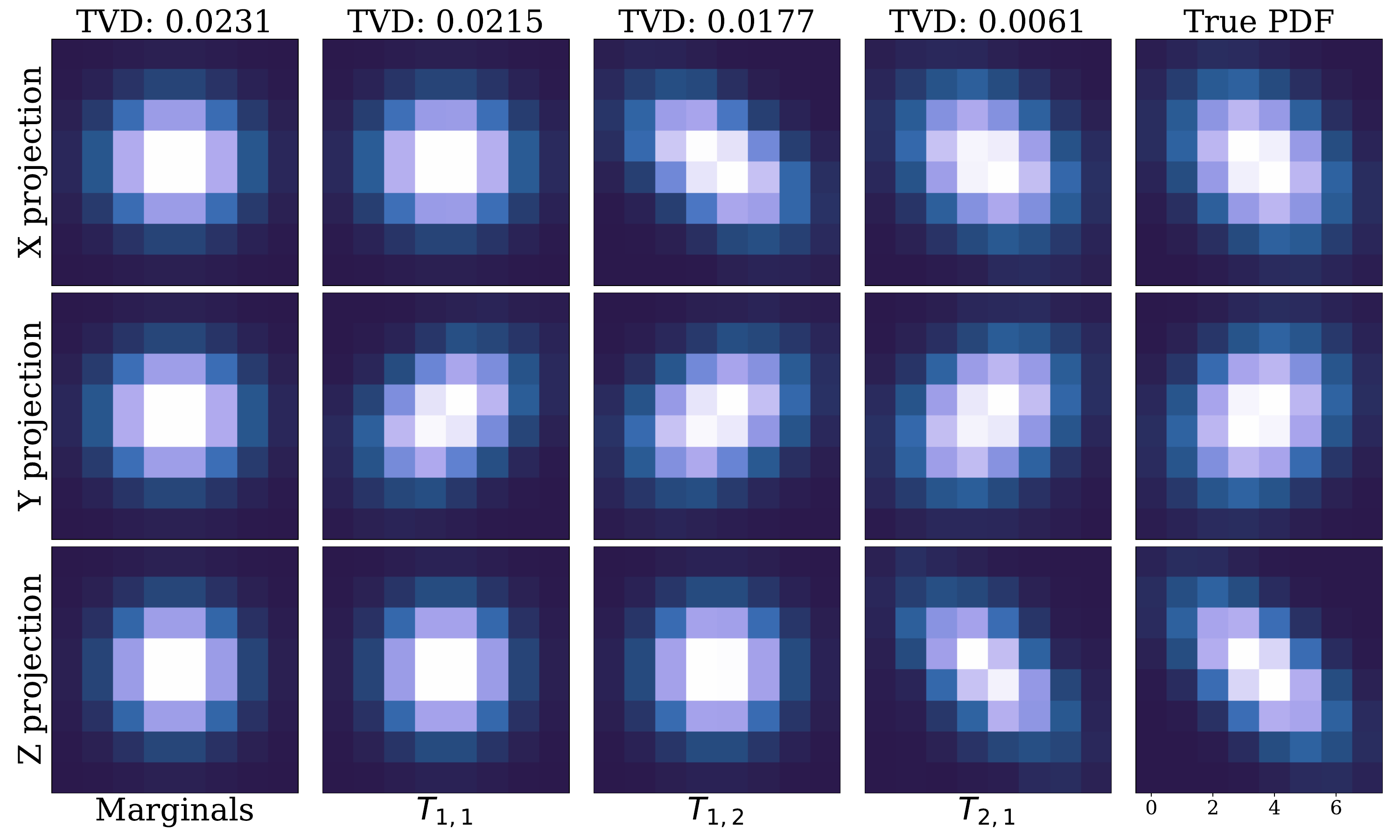}
\caption{\textbf{3D Correlated Gaussian}}
\label{fig:gaussian_3D_cubic}
\end{subfigure} & \begin{subfigure}[t]{0.96\columnwidth}
\centering 
\includegraphics[width=0.96\columnwidth]{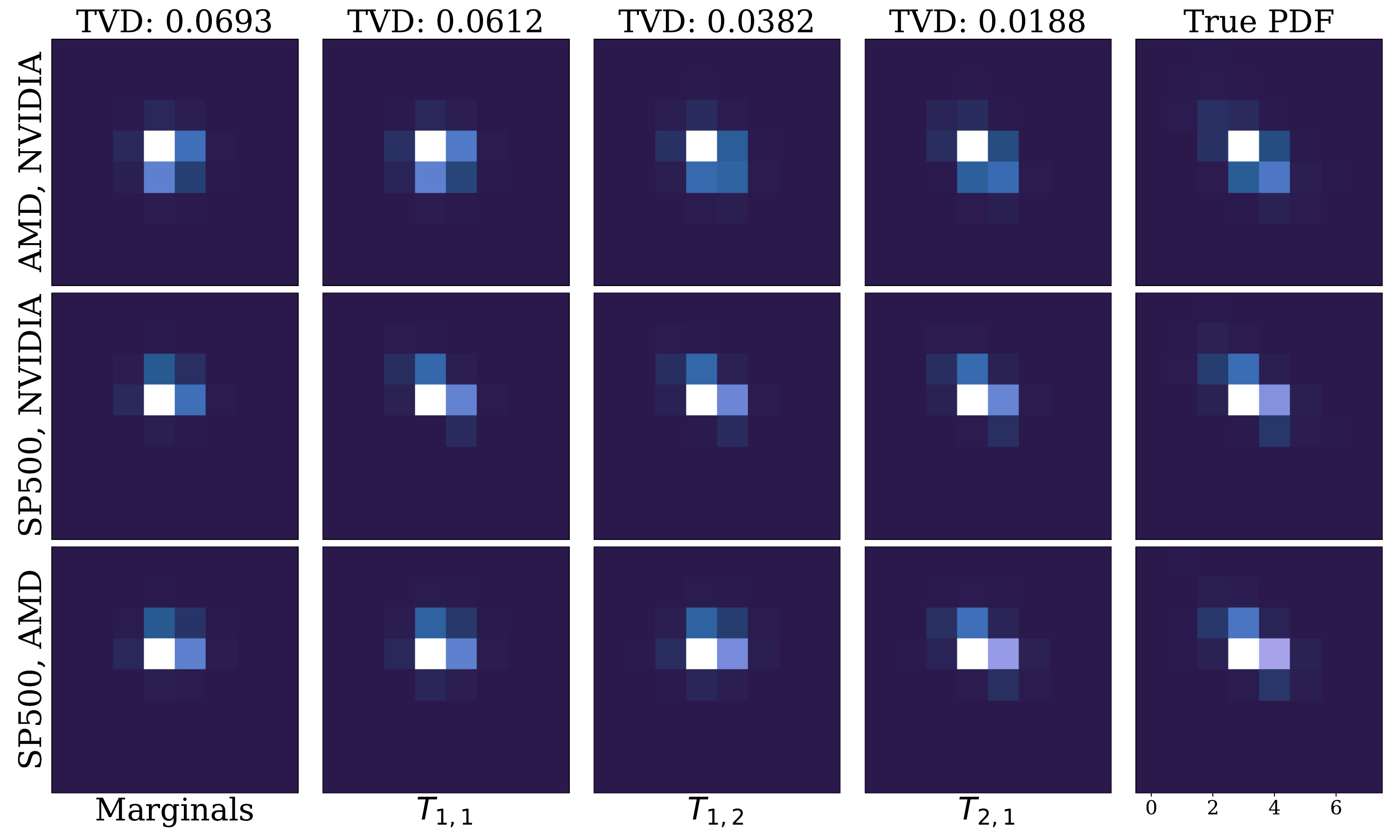}
\caption{\textbf{3 Asset Log-Return Distribution}}
\label{fig:stocks_3D_cubic}
\end{subfigure}
\end{tabular}
\caption{\textbf{Best tested number of layers to achieve convergence on 3D Gaussian and 3D stock price return distribution.} Progressive steps grow to the right, while the vertical axis shows projection on $X$, $Y$ and $Z$ axes, respectively. The last column shows the true projected distributions.}

\vspace{0.5em}

\begin{tabular}{c c}
\begin{subfigure}[t]{0.96\columnwidth}
\centering 
\includegraphics[width=0.96\columnwidth]{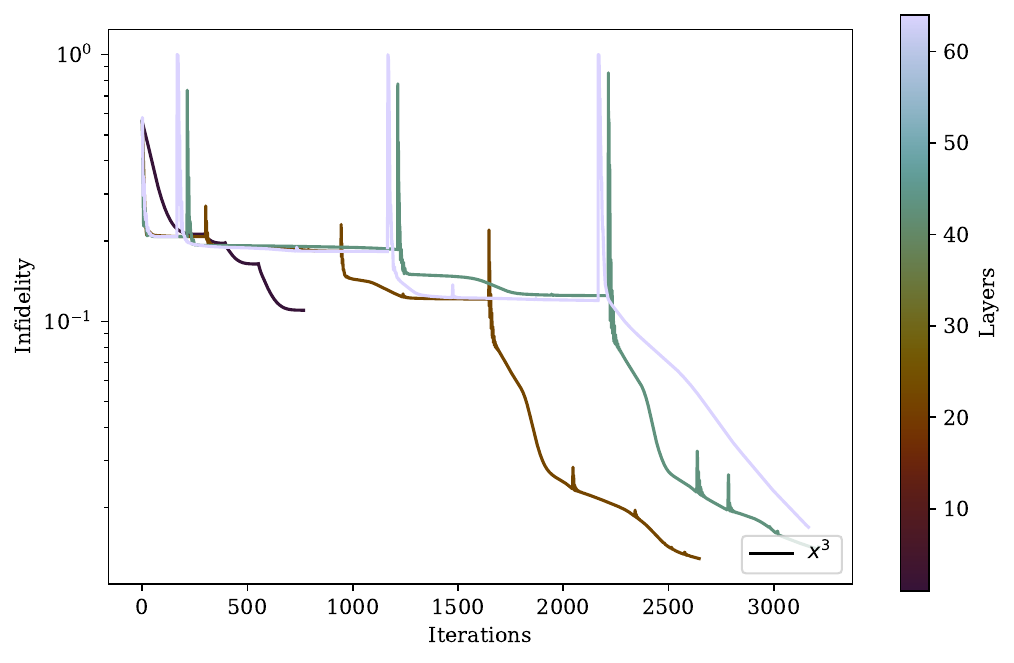}
\caption{\textbf{3D Correlated Gaussian}}
\label{fig:gaussian_3D_scaling_cubic}
\end{subfigure} & \begin{subfigure}[t]{0.96\columnwidth}
\centering 
\includegraphics[width=0.96\columnwidth]{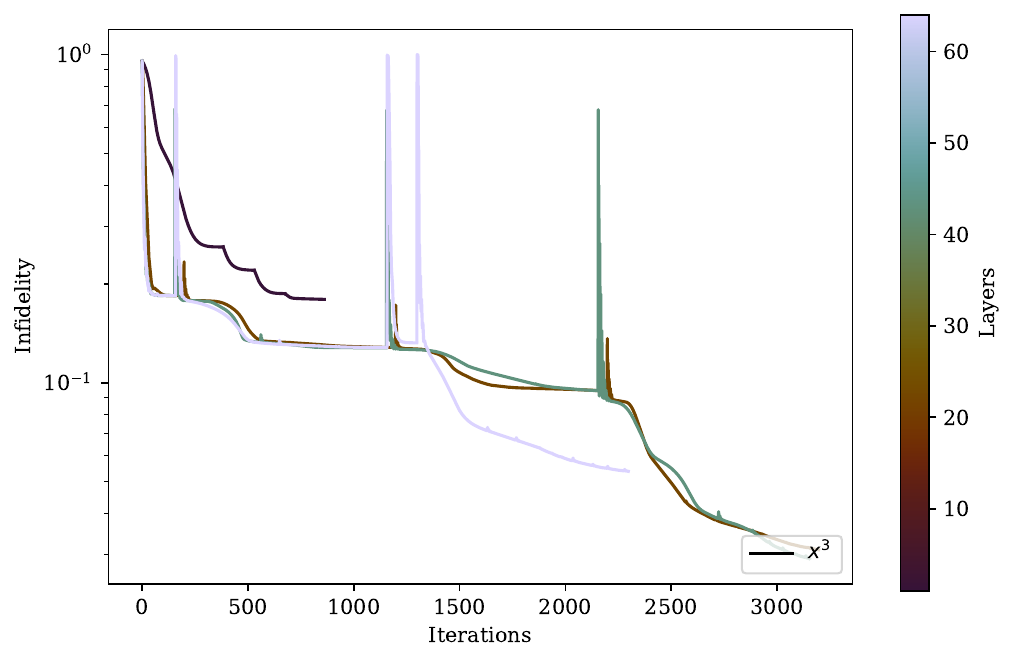}
\caption{\textbf{3 Asset Log-Return Distribution}} 
\label{fig:stocks_3D_scaling_cubic}
\end{subfigure} 
\end{tabular}
\caption{\textbf{Ablation study on the number of ansatz layers for achieving the best TVD with cubic scaling on 3D Gaussian and 3D stock price return distribution.} Progressive steps grow to the right, and the number of univariate and bivariate ansatz layers are illustrated via colored lines.}
\end{figure*}

\cref{fig:stocks_3D_cubic} and \cref{fig:stocks_4D_fixed} show the pdf comparisons when we fix the number of layers to $27$ and $9$ respectively. We perform an ablation study on the number of ansatz layers to converge on linear, quadratic and cubic scaling, and report cubic scaling for $3$ assets in \cref{fig:stocks_3D_scaling_cubic}. %

\begin{figure}[!t]
\centering 
\includegraphics[width=1.0\columnwidth]{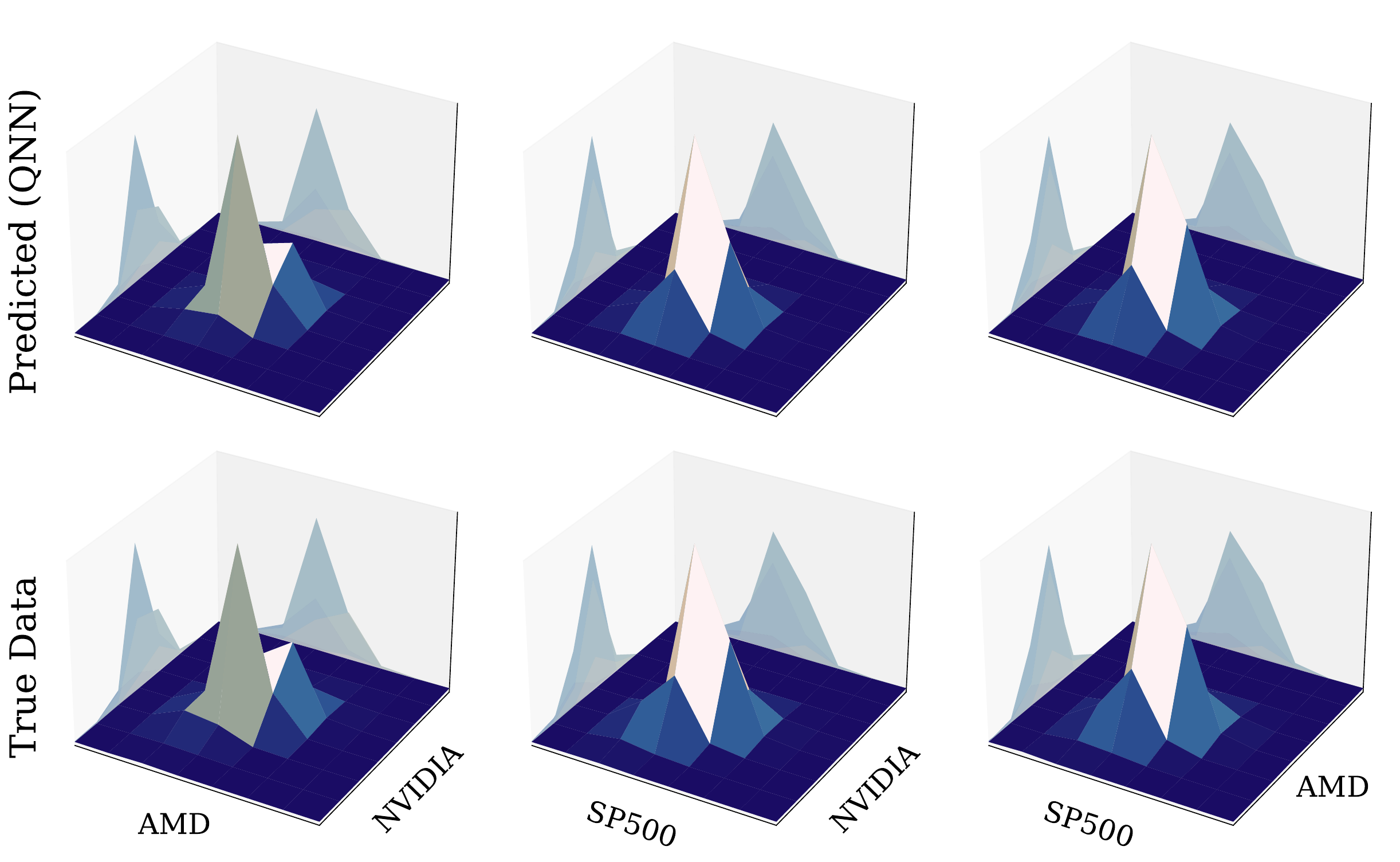}
\caption{\textbf{Projections of the best tested number of layers to achieve convergence on a 3D stock price return distribution.} The top row shows the trained pdf and the bottom row shows the original pdf. \cref{fig:stocks_3D_cubic} depicts throughout the progressive vine training.}
\label{fig:stocks_3D_projection}
\end{figure}

\section{Conclusion}

We present Qvine, a vine structured circuit ansatz and progressive training paradigm, that achieves high quality approximate loading of multi-dimensional distributions with highly scalable quantum circuits.

Given a vine structure, we define a circuit as depicted \cref{fig:vine_circ} and train the associated circuit blocks as shown in Alg.~\ref{alg:progtrain}. \cref{tab:resources} details the resources used for a $d$-dimensional distributions; with the number of one-qubit and two-qubit gates scaling as $\Oc\hm\left( d^2 \right)$, the depth of an R-vine bounded by $\Oc\hm\left( d^2 \right)$ while the depth of D-vines or other path-dominated decompositions scaling linearly $\Oc\hm\left( d \right)$. We test our vine ansatz on $3$-dimensional and $4$-dimensional gaussian as well as empirical price return distributions, showing high quality distribution loading through our ansatz. The optimal choice of $k$ for discretizing the domain of a univariate random variable, such as in the context of financial assets, is interesting to explore; see, for example, \cite{zhang2025asymptotic}.

\bibliographystyle{IEEEtran}
\bibliography{reference}

\end{document}